\documentstyle[prd,floats,aps]{revtex} \input psfig
\begin{document}
\title{The collision of boosted black holes: second order close limit
  calculations} \author{Carlos O. Nicasio$^{1}$, Reinaldo J.
  Gleiser$^1$, Richard H. Price$^2$,
  Jorge Pullin$^3$\\
  1. {\it Facultad de Matem\'atica, Astronom\'{\i}a y F\'{\i}sica,
    Universidad Nacional de C\'ordoba,\\ Ciudad
    Universitaria, 5000 C\'ordoba, Argentina.}\\
  2. {\it Department of Physics, University of Utah, Salt Lake City,
    Utah 84112.}\\
  3. {\it Center for Gravitational Physics and Geometry, Department of
    Physics\\The Pennsylvania State University, 104 Davey Lab,
    University    Park, PA 16802.}}

\maketitle
\begin{abstract}
  We study the head-on collision of black holes starting from
  unsymmetrized, Brill--Lindquist type data for black holes with
  non-vanishing initial linear momentum. Evolution of the initial data
  is carried out with the ``close limit approximation,'' in which
  small initial separation and momentum are assumed, and second-order
  perturbation theory is used.  We find agreement that is remarkably
  good, and that in some ways improves with increasing momentum.  This
  work extends a previous study in which second order perturbation
  calculations were used for momentarily stationary initial data, and
  another study in which linearized perturbation theory was used for
  initially moving holes.  In addition to supplying answers about the
  collisions, the present work has revealed several subtle points
  about the use of higher order perturbation theory, points that did
  not arise in the previous studies.  These points include issues of
  normalization, and of comparison with numerical simulations, and
  will be important to subsequent applications of approximation
  methods for collisions.
\end{abstract}
\vspace{-9.5cm} 
\begin{flushright}
\baselineskip=15pt
CGPG-98/2-1  \\
gr-qc/9802063\\
\end{flushright}
\vspace{8.5cm}

\section{Introduction}

The accurate prediction of gravitational waveforms produced in the
collisions of black holes has become a central topic of research in
general relativity, due to their potential observability with modern
interferometric gravitational wave detectors. Given the lack of
symmetries in a collision, it was believed for a long time that only a
full numerical integration of the Einstein equations would lead to
reliable answers. Recently it has been noticed \cite{PrPu,Pu} that one
can make some progress in understanding the collisions by using black
hole perturbation theory, especially for collisions in which the holes
start sufficiently close to each other for the collision to be
considered to be the evolution of a single, distorted black hole.
Following this approach, called the ``close limit approximation,''
linearized perturbation theory has been shown to provide a remarkably
accurate picture of the head-on collision of momentarily stationary
\cite{PrPu,Pu} and boosted black holes \cite{etaletal}.

If this technique is going to be considered a valid method for cases
in which full numerical simulations are still not available, one needs
to develop indicators for deciding when the approximation is
trustworthy.  Intuitive rules of thumb, such as requiring that a
single, almost spherical, horizon initially surround both holes turn
out to be too conservative to be practical, as was demonstrated by the
momentarily-stationary head-on collision results \cite{PrPu}. Recently
\cite{GlNiPrPuprl}, it was suggested that the use of second order
perturbation theory to provide ``error bars'' could be an effective
way of estimating the domain of validity of the first order results.
For the head-on collision of momentarily stationary black holes the
proposal appeared to work very well \cite{GlNiPrPuprl}.

The purpose of this paper is to explore the application of second order
calculations to an important case of initial data that is not
momentarily stationary, the head on collision of initially ``boosted''
(i.e., moving) holes. The introduction of boost turns out to add
several technical and conceptual complications. These are important
beyond their relevance to the specific collision studied here, since
the cases of realistic physical interest (collisions with spin and net
angular momentum) all involve initial data that is not momentarily
stationary. This paper will therefore attempt to lay part of the
groundwork for future investigations of more realistic situations
\cite{BaNo,BaNi}.

One of the conclusions of paper will be that perturbative calculations
can be a very reliable tool to get quantitative predictions at a
certain level of accuracy. If we wish to push the accuracy to a few
percent level, some questions remain. This is in part due to the fact
that numerical codes we use for comparison cannot at present be
trusted to that level of accuracy either, and in part due to several
technical complications that appear in perturbation theory. It is
remarkable, however, that the addition of considerable amounts of
boost to the black holes does not preclude the applicability of
perturbation theory techniques.

This paper is closely related to two previous studies to which
reference will frequently be made. The first is the linearized
analysis for initially boosted holes, by Baker {\it et
  al.}\cite{etaletal}; we shall refer to this as BAABRPS. The present
work relies heavily on the second order formalism described in
Refs.\,(\cite{GlNiPrPucqg,GlNiPrPuprl,GlNiPrPuprd}), which we shall
refer to collectively as GNPP.

In the present paper, Sec.\ II gives the details of how our initial
data is parameterized with a separation $L$ between the holes, and a
momentum parameter $P$. A discussion is given of the meaning of
perturbation theory in a two parameter space of solutions. It is also
pointed out that a feature of our initial data differs from that used
in computations with numerical relativity, and this difference hinders
a perfect comparison of results. We find, in this section, that to use
the second order formalism of GNPP, we must make gauge transformations
that eliminate the first order monopole perturbation of the extrinsic
curvature. This is the first of several technical issues that were not
evident in the linearized work of BAABRPS or the momentarily
stationary data of GNPP.

Section III shows how the initial data is evolved with a wave equation
that has the structure of a Zerilli\cite{zerilli} equation with a
source term quadratic in the first order perturbations. From these
results, it is shown how one computes the gravitational waveforms and
energies correct to second order, and the second-order correct results
are presented and are compared with results from numerical relativity.
In this section a discussion is given of a second order technical
detail that has previously been ignored and that was of little
consequence in the evolution of momentarily stationary initial data.
The gauge fixing used in GNPP leaves unfixed a degree of freedom
associated with time translations.  This is not relevant to the
computation of radiated energy, but must be resolved if our waveforms
are to be compared with those of numerical relativity. In Sec.\,III a
convenient method is given for fixing this gauge freedom in the
waveforms. With this choice, the second order correct waveforms as
well as energies are found to be in excellent agreement with the
results of numerical relativity.

The methods and results of the paper are briefly reviewed in Sec.\,IV
and the connection to future work is pointed out.

We use for the most part notation introduced in BAABRPS and GNPP,
which in turn is based on the notation of Regge and
Wheeler\cite{ReWh}. In addition, we will use here the convention of
adding superscripts in parentheses {\it when necessary} to indicate
whether a quantity is first or second order, and to indicate what
multipole it refers to. Multipole indices will be distinguished with
``$\ell=$.'' Thus, for example, $H^{(3)(\ell=2)}_0$ would indicate the
third order quadrupole Regge-Wheeler perturbation $H_0$.

\section{Initial data}

\subsection{The conformal approach}

The momentarily-stationary Misner \cite{Mi} initial solution to the
initial value problem of general relativity for a two black hole
situation has a convenient explicit analytical form; for initially
moving holes no such form is available and the first step in the
problem is to find an appropriate initial value solution.  We use the
conformal approach (see \cite{BoYo} and references therein), in which
one assumes the metric to be conformally flat $g_{ab} =\phi^4
\delta_{ab}$ and constructs the conformal extrinsic curvature
$\hat{K}_{ab} = \phi^{2} K_{ab}$. In terms of these variables the
initial value constraint equations (assuming maximal slicing ${\rm Tr}
K =0$) read,
\begin{eqnarray}
\nabla_a \hat{K}^{ab} &=& 0\label{momcons}\\
\nabla^2 \phi &=& -{1\over 8} {\hat{K}^{ab} \hat{K}_{ab} \over \phi^{7}}
\label{Hamil}
\end{eqnarray}
where all the derivatives are with respect to flat space.

One can construct \cite{BoYo} solutions to the first set of equations
(momentum constraint)  for a
single black hole  with linear momentum,
\begin{equation}\label{onehole}
\hat{K}^{\rm one}_{ab} = {3 \over  2 r^2} \left[ 2 P_{(a} n_{b)} -(\delta_{ab}
-n_a n_b)P^c n_c\right]\ .  \label{boyok}
\end{equation}
Here $r$ is the distance, in the conformally related flat space, from
the origin and $P_a$ is a vector in that space and can be shown to be
the momentum of the hole in the asymptotically flat physical space. By
superposing two such solutions one obtains the conformally related
extrinsic curvature representing two moving black holes (although the
flat space vectors $P_a$ in this case must be considered parameters
that have a clear interpretation as momentum only in the case that the
holes are widely separated).  Since the constraint equation
(\ref{momcons}) for the conformally related extrinsic curvature is
linear, the superposition still solves the constraint.  As was done in
BAABRPS, we locate the two black holes on the $z$ axis of the
conformally related flat space, at positions $z=\pm L$ and we choose
the flat space vectors $P_a$ to be symmetrically directed towards the
origin and to have equal size $P$. (The case of holes moving away from
the origin is represented with a negative value of $P$.)

As in BAABRPS, we treat the separation parameter $L$ as our
perturbation parameter, and we expand the two-hole superposed
$\hat{K}_{ab}$ in $L$. Due to the equal mass/opposite momentum
symmetry, only odd powers of $L$ appear, and the first two terms are:
\begin{eqnarray}\label{Kapprox}
\hat{K}_{ab} &=& {3 P L \over 2R^3} \left[
\begin{array}{ccc} -4 \cos^2 \theta&0&0\\0&R^2 (1+\cos^2 \theta)&0\\
0&0&R^2 \sin^2 \theta (3 \cos^2 \theta -1)
\end{array}\right]\\
&+&{3 P L^3 \over 16 R^5}
\left[
\begin{array}{ccc} 2 (1-18 \cos^2 \theta + 25 \cos^4 \theta)
&4 R \sin \theta \cos \theta (-1+5 \cos^2 \theta)&0\\
4 R \sin \theta \cos \theta (-1+5 \cos^2 \theta)
&R^2 (1+6\cos^2 \theta-15\cos^4 \theta)&0\\
0&0&R^2 (-3+33 \cos^2 \theta -65 \cos^4 \theta +35 \cos^6 \theta)
\end{array}\right]\nonumber.
\end{eqnarray}
Here $R$ is the flat space distance to the origin, related to the flat
space distance $d$ to the holes by $d=\sqrt{(R^2\pm
  2R\cos\theta+L^2)}$, and the expressions in Eq.\,(\ref{Kapprox}) are
valid only for $R> L$.

\subsection{Boundary conditions}

One must now put this solution in the right hand side of the
Hamiltonian constraint (\ref{Hamil}) and solve the resulting nonlinear
elliptic equation. In this process one needs to decide which boundary
conditions to impose for the elliptic problem. A common choice in
numerical studies has been the use of symmetrization of the data
through the two throats of the black holes (see for instance
\cite{Cook} and references therein). This kind of procedure is not
very convenient if one is interested in semi-analytic work as we are,
chiefly because symmetrizing implies using the method of images an
infinite number of times and the expressions involved become quite
large and difficult to handle. On the other hand, nothing prevents one
from constructing unsymmetrized data for boosted black holes along the
same lines as for the momentarily-stationary case (the
Brill--Lindquist problem \cite{BrLi}). This was recently emphasized by
Br\"ugmann and Brandt \cite{BrBr}. Here we will take this latter
approach. This is not completely inconsequential, since the only
numerical simulations available for comparison are for symmetrized
data; we will return to this point later. To generalize the
Brill--Lindquist construction to the case with momentum, one assumes
the conformal factor to be composed of two pieces,
\begin{equation}\label{twopiece}
\phi =  \phi_{\rm reg} +\phi_{BL}.
\end{equation}
One piece
\begin{equation}\label{BLpiece}
\phi_{BL} \equiv{m\over|\vec{R}-\vec{R}_1|} +{m\over|\vec{R}-\vec{R}_2|},
\end{equation}
is singular at the points $\vec{R}=\vec{R}_{1,2}$ in flat space, and
represents throats. That is, when one introduces a new radial coordinate
of the form $1/|\vec{R}-\vec{R}_{1,2}|$ the ``singular point ''
$\vec{R}=\vec{R}_{1,2}$ of the conformally related flat space is seen
to have the actual geometry of space that is asymptotically flat as
$|\vec{R}-\vec{R}_{1,2}|\rightarrow0$. The result of putting Eqs.\,
(\ref{twopiece}) and (\ref{BLpiece}) in (\ref{Hamil}) is
\begin{equation}\label{hamforphireg}
\nabla^2 \phi_{\rm reg} = -{1\over 8} {\hat{K}^{ab} \hat{K}_{ab}
\over (\phi_{BL}+\phi_{\rm reg})^{7}}\ ,\label{eqbrbr}
\end{equation}
which is to be solved for $\phi_{\rm reg}$ with the boundary
conditions that $\phi_{\rm reg}$ is regular at $\vec{R}=\vec{R}_{1,2}$
and approaches unity as $r\rightarrow\infty$.  Notice that the right
hand side of Eq.\,(\ref{hamforphireg}) is well behaved at
$\vec{R}=\vec{R}_{1,2}$; although the numerator diverges as
$|\vec{R}-\vec{R}_{1,2}|^3$, the denominator increases as
$|\vec{R}-\vec{R}_{1,2}|^7$.  The main difference between our approach
and that of Brandt and Br\"ugmann \cite{BrBr} is that we shall solve
the initial value problem perturbatively.

\subsection{Perturbative issues}

We have defined a two parameter ($P$ and $L$) family of initial data
that we could evolve into a two parameter family of spacetimes. We
are, of course, primarily interested in the close limit, the limit of
small initial separation, and hence of small $L$. In principle, we
could use initial data correct to second order in $L$.  This would
mean solving Eq.\,(\ref{hamforphireg}) for $\phi_{\rm reg}$ and
expanding the solution in $L$. In practice, Eq.\,(\ref{hamforphireg})
would require numerical solution since the right hand side of
Eq.\,(\ref{hamforphireg}) is regular, and the Green function for the
equation is simple, this would present no significant obstacle, but it
would have the disadvantage that we would have the solution only
numerically.  In particular, this would mean that the dependence on
$P$ would not be transparent.  For that reason, we follow a different
path.  As in BAABRPS, we consider small $P$ and small $L$. More
specifically, we consider a $P=\eta L$ curve in the family of spacetimes,
with $\eta$ a numerical factor of order one.
This means, for example, that terms proportional to $P^2,PL$ and $L^2$
are all of the same order, and are our lowest order perturbations.
Our second order perturbations will be of the form $P^4, P^3L,
\cdots$.  Due to the symmetry of our configuration, no terms arise of
order $P^3, P^2L,\cdots$.

Since $\hat{K}_{ab}$ has a leading factor of $P$, the numerator on
the right hand side of (\ref{Hamil}) is proportional to $P^2$. This
point is rather subtle, and there is a temptation to come to a wrong
conclusion. The expressions in (\ref{Kapprox}) are ${\cal O}(PL)$ and
suggest that the numerator in (\ref{Hamil}) is ${\cal O}(P^2L^2)$. It
must be understood, however, that the expressions in (\ref{Kapprox})
are valid only for $R>L$. The solution of (\ref{Hamil}) does not
depend locally only on the large $R$ form of the right hand side. It
connects boundary conditions at infinity with boundary conditions of
the throats, boundary conditions for which the $R>L$ condition does
not hold. Due to this non-locality in (\ref{Hamil}), or equivalently
(\ref{hamforphireg}), the conformal factor depends in a complicated,
nonpolynomial, way on the $P$ parameter. Our ``small $P$'' assumption
amounts to taking the numerator in (\ref{Hamil}), or equivalently
(\ref{hamforphireg}), to be perturbative, at all points in space.
This, and closely related issues, are further discussed in BAABRPS.

The perturbative problem requires at several points the specification
of a ``mass'', either of the spacetime or of the black holes. Let us
discuss this in some detail. First, there is the problem of what mass
does one use for the background Schwarzschild spacetime around which
we are doing perturbation theory. Our experience shows that one should
use the ADM mass of the spacetime. We saw a similar situation when we
analyzed the radiation generated as an initially conformally flat
(``Bowen-York''\cite{BoYo}) spinning hole\cite{GlNiPrPuby} settles
into its Kerr final state.  The spin rate was taken to be small, and
the problem was treated as a perturbation away from the Schwarzschild
geometry.  For apparently moderate amounts of spin, the radiation
generated was rather small, but the effect on the ADM mass (i.e., the
spin dependent increase over the Schwarzschild mass) could be a factor
of several.  By computing exactly the effect of spin on the ADM mass,
we found we could successfully apply perturbation theory for moderate
spin.The only question could be if one uses that of the initial slice
or that after the radiation has gone out, but for the cases of
interest the difference is less than $1\%$, so we will consider the
ADM mass of the initial slice for our background. 

Then there is the issue of the initial data. The initial data for
boosted black holes is characterized by the separation $L$, the
momentum $P$ and a ``bare'' mass for each hole $m$, which also serves
as overall scale factor.  This mass has no clear physical meaning, and
no equivalent in the reflection symmetric initial data used in
numerical relativity.  Because of this, we would prefer to have the
initial data parameterized by $P, L, M_{ADM}$.  Since $P, L$ and $m$
determine uniquely the ADM mass, this is formally no problem. In
practice we proceed in the following way. The ADM mass (for a given
set of parameters $L,P,m$) can be found from the monopole part of our
second order solution for the conformal factor \cite{Gl}.  One can
then write an expansion for $m$ of the form,
\begin{equation}
2 m = M_{ADM} + P^2 L^2 m_1 +\ldots \label{expanmadm}
\end{equation}
with $m_1$ a constant. One can then take the intial data, $g_0(P,L,m)$
and rewrite it as $g_0(P,L,M_{ADM}(P,L,m))$ and use the above
expansion (\ref{expanmadm}) for the explicit form of
$M_{ADM}(P,L,m)$. As a result of this reparameterization, the first
and second order terms of the initial data are left invariant. That
is, one simply takes the initial data and where it read ``$2m$'' one
replaces $M_{ADM}$.  For the second order pieces this is also true,
the second order pieces of (\ref{expanmadm}) only contribute
irrelevant $\ell=0$ terms to second order and do not change the
initial data. Summarizing, we construct the perturbative initial data,
and wherever it said $m$ we replace $M_{ADM}/2$ and this is consistent
to the order of perturbation theory we are considering. Therefore our
problem is completely parameterized now by the ADM mass, which also
facilitates comparison with the full numerical data, which are also
parameterized and normalized by the ADM mass. This issue was the
source of significant confusion in this area initially. In particular,
the results of \cite{etaletal} are not properly normalized and
therefore depart from our predictions in this paper for moderate and
large values of the momentum (when $M_{ADM}$ starts to differ
significantly from twice the ``bare mass'' of the holes).

The concrete details of computation start with (\ref{hamforphireg}).
For $R>L$ and for computations only to second order, one needs only
the portion linear in $L$ of the extrinsic curvature. Keeping terms
only to second order, and taking on the right hand side of the
Hamiltonian constraint, the form of $\phi=\phi_{BL}$ given in
(\ref{BLpiece}), one gets the $O(L^2 P^2)$ piece of the conformal
factor,
\begin{equation}
\phi_{\rm reg}\;=\; P^2 L^2 \phi^{(2)}\;+\;\mbox{higher order terms},
\end{equation}
by solving,
\begin{equation}
P^2L^2\nabla^2\phi^{(2)}\;=\; S^{(2)} \;=\;
-72 P^2 L^2 { R (1-2\cos^2\theta^2+13\cos^4\theta)\over 
(2 R+m)^7}
\label{eqconf2}
\end{equation}
with the boundary condition $\phi^{(2)}\rightarrow0$ at
$R\rightarrow\infty$.  We can simplify the solution of this Poisson
equation by decomposing the source into multipoles:
\begin{eqnarray}
S^{(2)(\ell=0)} &=& -{1056\over 5} {P^2 L^2 R\over(2 R+m)^7} \nonumber \\
S^{(2)(\ell=2)} &=& -{3072\over 7} {P^2 L^2 R\over(2 R+m)^7} \nonumber \\
S^{(2)(\ell=4)} &=& -{7488\over 35} {P^2 L^2 R\over(2 R+m)^7}\ . \nonumber
\end{eqnarray}
The solution for the monopole and quadrupole parts are:
\begin{eqnarray}
\phi^{(2)(\ell=0)} &=& -{11\over 50} {P^2 L^2 (8 R m
+20 R^2+m^2)\over (2 R+m)^5 R}\;+\;
{P^2 L^2 q_0\over R} \nonumber \\
\phi^{(2)(\ell=2)} &=& -{8\over 35} {P^2 L^2 (m^4
+10 m^3 R+40 m^2 R^2+80 m R^3+80 R^4)
\over (2 R+m)^5 R^3}\;+\;{P^2 L^2 q_2\over R^3} \label{phis}.
\end{eqnarray}
One can also solve for the $\ell=4$ piece but we will not need it.

The solution contains two constants $q_0$ and $q_2$ representing the
homogeneous solution of the Poisson equation. These constants
determine, in effect, what boundary conditions are being chosen for
the conformal factor. The choice we have made is that $\phi_{\rm
  reg}$, and hence $\phi^{(2)}$, is regular everywhere. The wrong
choice of $q_0$ or $q_2$ means that when the solutions in (\ref{phis})
are continued to $R<L$ they will be singular, so that $\phi=\phi_{BL}$
does not contain all the information about the singularities.  (This
would be the case, for instance, if we took $q_0$ and $q_2$ to have
the values for the symmetrized solution.)

To determine what values of $q_0$ and $q_2$ give a regular solution,
we start by noticing that from (\ref{phis}) we can see that the
asymptotic form of the conformal factor is,
\begin{eqnarray}
\phi^{(2)(\ell=0)} &=& {P^2 L^2 q_0\over R}\;+\;O(1/R^4)\nonumber \\
\phi^{(2)(\ell=2)} &=& {P^2 L^2 q_2\over R^3}\;+\;O(1/R^4) \nonumber \\
\phi^{(2)(\ell=4)} &=& O(1/R^4). \label{pertq}
\end{eqnarray}

On the other hand, we know that the regular part of the conformal
factor admits an expansion of the form,
\begin{equation}
\phi_{\rm reg}\;\simeq\; \sum_{n=1}^\infty {A_n(\theta,P,L)\over R^n}
\label{phifar}
\end{equation}
Since the right hand side of (\ref{hamforphireg}) falls off as $1/R^6$
one can conclude that the first three coefficients of the above
expansion are part of the homogeneous solution, and have the form
\begin{eqnarray}
A_0(\theta,P,L) &=& Q_0(P,L) \nonumber \\
A_1(\theta,P,L) &=& 0 \nonumber \\
A_2(\theta,P,L) &=& Q_2(P,L) P_2(\cos(\theta))\ . \label{exactq}
\end{eqnarray}
We therefore see that $q_0$ and $q_2$ are just the leading
coefficients in an expansion of $Q_0$ and $Q_2$ in terms of $P$ and
$L$.

One can obtain a closed form expression for $Q_0$ and $Q_2$ by
applying Gauss' theorem to (\ref{hamforphireg}), and using the fact
that (by choice) $\phi_{\rm reg}$ is regular on the whole $R,\theta$
plane $\Sigma^2$. This takes the form
\begin{equation}
\int_{\partial \Sigma^2} \vec{\nabla} \phi_{\rm reg} \cdot \vec{d\Sigma} = 
\int_{\Sigma^2} S(R,\theta,P,L)\ d^2S,
\end{equation}
where $S(R,\theta,P,L)$ represents the right hand side of
(\ref{hamforphireg}), and the integral on the left is evaluated over
the boundary of the plane at infinity. It is clear that the only term
that contributes to this integral is the leading term for the
expansion of $\phi$. From there we can therefore determine $Q_0$.
Considering the same construction, now for $R^2 \phi$ one can
determine $Q_2$. The results are,
\begin{eqnarray}
Q_0(P,L) &=& -{1\over 2} \int_0^\infty dr \int_0^\pi d\theta\;R^2 \sin(\theta) 
 P_0(\cos(\theta)) S(R,\theta,P,L) \label{qus} \\
Q_2(P,L) &=& -{1\over 2} \int_0^\infty dr \int_0^\pi d\theta\;R^4 \sin(\theta) 
 P_2(\cos(\theta)) S(R,\theta,P,L) \nonumber 
\end{eqnarray}
so therefore for the leading terms we get,
\begin{eqnarray}
q_0 &=& -{1\over 8} {\partial^4\over \partial P^2 \partial L^2} \left. \left[
\int_0^\infty dr \int_0^\pi d\theta\;R^2 \sin(\theta) 
 P_0(\cos(\theta)) S(R,\theta,P,L) \right] \right|_{P=0,L=0} \label{quspert}\\
q_2 &=& -{1\over 8} {\partial^4\over \partial P^2 \partial L^2} \left. \left[
\int_0^\infty dr \int_0^\pi d\theta\;R^4 \sin(\theta) 
 P_2(\cos(\theta)) S(R,\theta,P,L) \right] \right|_{P=0,L=0} \nonumber.
\end{eqnarray}

These expressions are straightforward to evaluate, especially since to
the order of interest we can replace the source by, 
\begin{equation}
S^{(P)}\;=\;-{1\over 8} {\widehat{K}^2 \over \phi_0^7}, 
\end{equation}
where $\widehat{K}^2$ is the square of the trace of (\ref{Kapprox}) and
$\phi_0$ is the conformal factor evaluated for $P=0$.  The integrals
(\ref{quspert}) however, cannot be solved in closed form. Instead they
were computed numerically (in several different ways).  The numerical
treatment of $S(R,\theta,P,L)$ requires some care. As pointed out near
the end of Sec.\,IIB, though the source term is regular, both the
numerator and denominator on the right hand side diverge at the points
representing the holes. The results we get for the constants are
\begin{equation}
q_0 = 0.219/m^3,\qquad
q_2 = 0.224/m.
\end{equation}
These numbers are in excellent agreement with an approximate
calculation due to Brandt and Br\"ugmann \cite{BrBr}. They obtain
approximately the correction of the ADM mass due to the momentum in
the initial data. The leading term in the expansion in $L$ of their
formula is precisely our $q_0$. Their result is $11/50\sim 0.22$.  One
can reproduce these formulas by considering expansions of the
integrals considered in powers of $L$.

\subsection{Casting the initial data in the Zerilli formalism}

Having the initial data for the problem, we now can input it into the
perturbation formalism and evolve it. The first order perturbations
are evolved with a Zerilli equation. The second order perturbations
are evolved with a Zerilli equation with a ``source'' term quadratic
in first order perturbations. The details of how this is done for the
momentarily stationary Misner\cite{Mi} initial data was described in
GNPP. Those details, however, were rather specific to the Misner case.
In particular, the formalism in that work used the fact that in the
Misner initial metric the only first order perturbations are
quadrupolar, and hence the source in the second order Zerilli equation
is constructed entirely from $\ell=2$ first order perturbations. Those
details also assumed that certain of the second order initial metric
perturbations vanished.

It will be convenient to use gauge transformations to
satisfy these same conditions, so that the previous formalism can be
used.  The initial metric (because it is conformally flat) has the
correct Misner-like second order form. The extrinsic curvature,
however, has a first order $\ell=0$ perturbation which generates
$\ell=0$ perturbations in the evolved data. These $\ell=0$
perturbations would contribute to the source term of the second order
Zerilli equation. Below we will use a first order gauge transformation
to eliminate this first order perturbation. This transformation,
however, changes the second order initial metric, taking it out of
``Misner form.''  We then use a second order gauge transformation to
restore it to the Misner form.

Let us start by writing the perturbations in the standard
Regge--Wheeler \cite{ReWh} notation for the multipolar decomposition
of a metric tensor $g_{ab}$, ie,
\begin{eqnarray}
g_{tt} &=& (1-2M/r)\sum_\ell H_0^{(\ell)}(r,t) 
P_\ell(\cos\theta)\label{firstRW}\\
g_{tr} &=& \sum_\ell H_1^{(\ell)}(r,t)P_\ell(\cos\theta)\\
g_{t\theta} &=& \sum_\ell h_0^{(\ell)}(r,t) 
(\partial/\partial\theta) P_\ell(\cos\theta)\\
g_{r\theta} &=& \sum_\ell h_1^{(\ell)}(r,t) 
(\partial/\partial\theta) P_\ell(\cos\theta)\\
g_{rr}&=& (1-2M/r)^{-1} \sum_\ell H_2^{(\ell)}(r,t) P_\ell(\cos\theta)\\
g_{\theta\theta} &=& r^2\sum_\ell\left[K^{(\ell)}(r,t)  +
G^{(\ell)}(r,t)(\partial^2/\partial\theta^2) \right]P_\ell(\cos\theta)\\
g_{\phi\phi} &=& r^2\sin^2\theta\sum_\ell\left[K^{(\ell)}(r,t)  +
G^{(\ell)}(r,t)\cot\theta(\partial/\partial\theta) 
\right]P_\ell(\cos\theta)\label{lastRW}\ .
\end{eqnarray}
In these expressions, $r$ is related to $R$, the radial coordinate in
the conformally related flat space, by $R = (\sqrt{r}
+\sqrt{r-2M})^2/4$. The ``background mass'' $M$, as previously
discussed, is the ADM mass computed numerically for a given choice of
$m,L,P$.

Since the initial geometry is conformally flat, the only non-vanishing
perturbations are those in $H_2$ and $K$. The quadrupole parts, to
second order, of these perturbations are:
\begin{eqnarray}
H^{(\ell=2)}_2&=& K^{(\ell=2)} = {16ML^2 \over \sqrt{r} (\sqrt{r}+\sqrt{r-2M})^5} + {192 M^2 L^4 \over 7 r (\sqrt{r}+\sqrt{r-2M})^{10}}+ {128 L^2 P^2 {\it q_2} \over \sqrt{r} (\sqrt{r}+\sqrt{r-2M})^5} \\
&&
-{256[12r^2-9rM+M^2 +(8r-3M)\sqrt{r}\sqrt{r-2M}] L^2 P^2 \over 35 r^3 (\sqrt{r}+\sqrt{r-2M})^6} \nonumber
 \end{eqnarray}

To describe the perturbations of the extrinsic curvature we shall use
a notation like that in (\ref{firstRW})--(\ref{lastRW}), but shall
prefix extrinsic curvature quantities with a ``$K$''. Thus, for
example, $K_{rr} =\sum_\ell(1-2M/r)K\!H_2^{(\ell)}P_\ell(\cos\theta)$.
The non-vanishing monopole perturbations of the extrinsic curvature are
\begin{eqnarray}
K\!H^{(\ell=0)}_2&=&2 {P L\over r^3}\\
K\!K^{(\ell=0)}&=& - {P L\over r^3}
\end{eqnarray}
and the quadrupole perturbations are
\begin{eqnarray}
K\!H^{(\ell=2)}_2&=&
{\frac {4\,PL}{{r}^{3}}}+{\frac {16\,P{L}^{3}\left (6\,r-11\,M+6\,
\sqrt {r}\sqrt {r-2\,M}\right )}{7\,{r}^{7/2}\left (\sqrt {r}+\sqrt {
r-2\,M}\right )^{5}}}\\
K\!G^{(\ell=2)} &=&
-{\frac {PL}{{r}^{3}}}+{\frac {8\,P{L}^{3} 
\sqrt {r-2\,M}}{7\,{r}^{7/2}\left (\sqrt {r}+\sqrt {r-2\,M}
\right )^{4}}}\\
K\!K^{(\ell=2)} &=&
-{\frac {5\,PL}{{r}^{3}}}-{\frac {8\,P{L}^{3}\left (3\,r-5\,M+3\,
\sqrt {r}\sqrt {r-2\,M}\right )}{7\,{r}^{7/2}\left (\sqrt {r}+\sqrt {
r-2\,M}\right )^{5}}}\\
K\!h^{(\ell=2)}_1 &=&
-{\frac {32\,P{L}^{3}}{7\,\sqrt {r-2\,M} {r}^{3/2}\left (\sqrt {r}+\sqrt {r-2\,M}\right )^{
4}}}\ .
\end{eqnarray}

We start the process of gauge transformations by writing 
a general $\ell=0$ and $\ell=2$ 
first order gauge transformation
vector,
\begin{eqnarray}
\xi^t&=& M^{(1)(\ell=2)}_0 P_2(\cos\theta) +M^{(1)(\ell=0)}_0\\
\xi^r &=& M^{(1)(\ell=2)}_1 P_2(\cos\theta)+M^{(1)(\ell=0)}_1\\
\xi^\theta &=& M^{(1)(\ell=2)}_a 
{\partial P_2(\cos\theta)\over \partial \theta}\ ,
\end{eqnarray}
and we choose all components to vanish except,
\begin{eqnarray}
M^{(1)(\ell=0)}_0 &=& {\sqrt{r} \over M \sqrt{r-2 M}} P L\label{firstM0}\\
M^{(1)(\ell=0)}_1 &=& -{\sqrt{r-2 M}\over r^{5/2}} t PL.\label{secondM0}
\end{eqnarray}
This gauge transformation eliminates the $\ell=0$ perturbation of the
extrinsic curvature to first order, and leaves the $\ell=2$ first
order initial data unchanged, but it introduces quadratic changes in
the second order components of the initial data.  To compute these
second order changes, we need a four dimensional metric, whereas up to
now we have only dealt with the initial values. We assume zero
perturbative lapse and shift to all orders and use the initial data to
write an expansion in powers of a fiducial time $t$ for the four
dimensional metric around $t=0$,
\begin{equation}
g_{\mu\nu} = (g_{\mu\nu})_{t=0} + t (\partial_t g_{\mu\nu})_{t=0}+\ldots
\end{equation} 
where $(g_{\mu\nu})_{t=0}$ is constructed in a straightforward manner
with the 3-metric and the chosen lapse and shift, and the time
derivative of the perturbative piece of the metric $h_{\mu\nu}$ is
completely determined by the extrinsic curvature,
\begin{equation}
(\partial_t h_{ab})_{t=0} = -2 K_{ab} \sqrt{1-{2M\over r}}.
\end{equation}
We then apply the formulas for gauge transformations to the above
constructed metric and  take the limit $t\rightarrow0$ to recover
the initial data in the new gauge.

The second order changes due to quadratic combinations of the first
order gauge transformation have $\ell=0$ and $\ell=2$ components.  We
will ignore the first, since they are non-radiative.  The $\ell=2$
second order metric that results from the gauge transformation of
(\ref{firstM0})--(\ref{secondM0}) is:
\begin{eqnarray}
h_1^{(2)(\ell=2)} &=&
{\frac {8\,P{L}^{3}t\left (r^{3/2}-3 M r^{1/2} -(r+2 M) \sqrt{r-2M}\right )}{M\sqrt{r-2\,M} {r}^{2}\left (\sqrt {r}
+\sqrt {r-2\,M}\right )^{4}}}\label{firstpost}\\
H_1^{(2)(\ell=2)} &=& 
-{\frac {16\,P{L}^{3}M}{{r}^{
2}\sqrt {r-2\,M}\left (\sqrt {r}+\sqrt {r-2\,M}\right )^{5}}}+{
\frac {8\,t{P}^{2}{L}^{2}}{{r}^{5}}}\\
H_2^{(2)(\ell=2)} &=& 
-{\frac {8\,{P}^{2}{L}^{2}}{M{r}^{3}}}
+{\frac {16\,P{L}^{3}t\left [r^{1/2} \left(r+4M\right) \sqrt {r-2\,
M}+3M^2+3rM-r^2 \right]} {M {r}^{4} \left(\sqrt{r}+
\sqrt {r-2\,M}\right)^{4}}}\\
K^{(2)(\ell=2)} &=&
{\frac {10\,{P}^{2}{L}^{2}}{M{r}^{3}}}-{\frac {8\, P 
{L}^{3}t \left (2M^2-5rM+2r^2-2 (r-3M) \sqrt{r} \sqrt{r-2M}
 \right )}{{r}^{4} M\left (\sqrt {r}+
\sqrt {r-2\,M}\right )^{4}}} \\
G^{(2)(\ell=2)} &=&
{\frac {2\,{P}^{2}{L}^{2}}{M{r}^{3}}}+{16PL^3t\sqrt{r-2M} \over r^3(\sqrt{r}+\sqrt{r-2M})^5}
\label{lastpost}.
\end{eqnarray}

In the formalism of GNPP the initial data was taken to have
$H_0=H_1=h_0=0$ up to second order. This was true of our perturbed
metric before the gauge transformation of
(\ref{firstM0})--(\ref{secondM0}), but is not true of the
post-transformation metric of (\ref{firstpost})--(\ref{lastpost}). We
now restore the conditions $H_0=H_1=h_0=0$, for the quadrupole, with
another, purely second order, gauge transformation:
\begin{eqnarray}
M^{(\ell=2)}_0 &=&{\displaystyle \frac {4}{3}} t^{2}\,P\,L^{2}\,M
{ - 6\,L
\,r^{3}M + P t\,\sqrt{r - 2\,M} \,(\sqrt{r}+\sqrt{r-2M})^5 
 \over r^{7}\sqrt{ r-2M}(\sqrt{r}+\sqrt{r-2M})^5}\\
M^{(\ell=2)}_1 &=& 4t\,P\,L^{2}\,\sqrt{r-2M}\left[
{ 4 L\,r^{3}M -Pt\sqrt{r - 2\,M
}  (\sqrt{r - 2\,M}+\sqrt{r})^5 \over r^{6}(\sqrt{r}+\sqrt{r-2M})^5  }\right]\\
M^{(\ell=2)}_a &=&{\displaystyle \frac {1}{3}} P\,L^{2}\,M \,t
^{3} 
{-8 L\,r^{3} M\sqrt{r - 2\,M}+ P\,t\,(r - 2\,M)(\sqrt{r}+\sqrt{r-2M})^5    \over r^{10}(\sqrt{r}+\sqrt{r-2M})^5}.
\end{eqnarray}

With this transformation, the final form of the first and second order
parts of the quadrupole metric perturbations read,
\begin{eqnarray}
H_2^{(1)(\ell=2)} &=& K^{(1)(\ell=2)} =
 16\,{\displaystyle \frac {M\,L^{2}}{
(\sqrt{r} + \sqrt{r - 2\,M})^{5}\,\sqrt{r}}} \label{final3geom}\\
H_2^{(2)(\ell=2)} &=&
-{1\over 35} {L}^{2}{P}^{2}\left[ 1272\,{r}^{5/2}+1240\,\sqrt {r-2\,M}{r}^{2}+2480\,
\left (r-2\,M\right ){r}^{3/2}+2480\,\left (r-2\,M\right )^{3/2}r \right. +
\nonumber \\
&+&\left. 1240\,\left (r-2\,M\right )^{2}\sqrt {r}+248\,\left (r-2\,M\right )^{5/2} \right] \left/ \left[ 
\left(\sqrt {r}+\sqrt {r-2\,M}\right )^{5}{r}^{3}M \right] \right. + \nonumber \\
&+& {\frac {128\,{L}^{2}{P}^{2}{\it q2}}{\left (\sqrt {r}+\sqrt {r-2\,M}
\right )^{5}\sqrt {r}}}+{\frac {192\,{M}^{2}{L}^{4}}{7\,\left (\sqrt {
r}+\sqrt {r-2\,M}\right )^{10}r}} \nonumber \\
K^{(2)(\ell=2)} &=&
- {1\over 35} {L}^{2}{P}^{2} \left[ 642\,{r}^{5/2}-1910\,\sqrt {r-2\,M}{r}^{2}-3820\,
\left (r-2\,M\right ){r}^{3/2}-3820\,\left (r-2\,M\right )^{3/2}r - \right. \nonumber \\
&-&\left. 1910 \,\left (r-2\,M\right )^{2}\sqrt {r}-382\,\left (r-2\,M\right )^{5/2} \right] \left/ \left[ \left (\sqrt {r}+\sqrt {r-2\,M}\right )^{5
}{r}^{3}M \right] \right. \nonumber \\ 
&+& {\frac {128\,{L}^{2}{P}^{2}{\it q2}}{\left (\sqrt {r}+\sqrt {r-2\,M}
\right )^{5}\sqrt {r}}}+{\frac {192\,{M}^{2}{L}^{4}}{7\,\left (\sqrt {
r}+\sqrt {r-2\,M}\right )^{10}r}} \nonumber \\
G^{(2)(\ell=2)} &=&{\frac {2\,{P}^{2}{L}^{2}}{M{r}^{3}}} \nonumber 
\end{eqnarray}

and the extrinsic curvature is,
\begin{eqnarray}\label{finalext}
K\!H_2^{(1)(\ell=2)} &=& 4 P L/r^3\\
K\!G^{(1)(\ell=2)} &=& -P L/r^3 \nonumber  \\
K\!K^{(1)(\ell=2)} &=& -5P L/r^3 \nonumber \\
K\!h_1^{(2)(\ell=2)} &=&
{\frac {\left (108\,\sqrt {r}+52\,\sqrt {r-2\,M}\right )P{L}^{3}}{7\,
\left (\sqrt {r}+\sqrt {r-2\,M}\right )^{5}{r}^{3/2}\sqrt {r-2\,M}}} \nonumber \\
K\!H_2^{(2)(\ell=2)} &=&
-{\frac {\left (296\,r+184\,\sqrt {r}\sqrt {r-2\,M}+64\,M\right )P{L}^
{3}}{7\,\left (\sqrt {r}+\sqrt {r-2\,M}\right )^{5}{r}^{7/2}}} \nonumber \\
K\!G^{(2)(\ell=2)} &=& 
-{\frac {\left (48\,r-8\,\sqrt {r}\sqrt {r-2\,M}+16\,M\right )P{L}^{3}
}{7\,\left (\sqrt {r}+\sqrt {r-2\,M}\right )^{5}{r}^{7/2}}} \nonumber \\
K\!K^{(2)} &=& 
{\frac {\left (4\,r+116\,\sqrt {r}\sqrt {r-2\,M}-16\,M\right )P{L}^{3}
}{7\,\left (\sqrt {r}+\sqrt {r-2\,M}\right )^{5}{r}^{7/2}}} \nonumber
\end{eqnarray}

For perturbations satisfying the Misner conditions ($H_0=H_1=h_0=0$)
the first order, quadrupole, Zerilli function, in the notation, 
of GNPP is given by
\begin{equation}
\psi = {r-2M \over 3(2r+3M)}\left[ r H_2^{(1)(\ell=2)}+3r^2G^{(1)(\ell=2)}_r -
 r^2 K^{(1)(\ell=2)}_r  -6 h_1^{(1)(\ell=2)} \right] + { r \over 3} K^{(1)(\ell=2)} 
 \end{equation}
and its time derivative by,
\begin{equation}
\partial_t \psi = 
 {2\,\sqrt {r-2\,M} \over \sqrt {r}\left (2\,r+3\,M\right )}
\left[- r^{2} KK^{(1)(\ell=2)}_{} - r\left(
3\,M-r\right ) KG^{(1)(\ell=2)}_{} + \left (r-2\,M\right )
Kh_1^{(1)(\ell=2)} \right]
\end{equation}

Here use has been made of the first order Einstein equations to
simplify the occurrence of higher time derivatives.

With the notation and formalism of GNPP, the second order, $\ell=2$
Zerilli function $\chi$ is computed to be 
\begin{eqnarray}
\chi &=&\left[
-14\,r^{4}\rho\,\left (-
 K\!G^{(2)}_{} r^{2}+r^{2} K\!K^{(2)}_{} +3\,r K\!G^{(2)}_{} M-
 K\!h_1^{(2)} r+2\, K\!h_1^{(2)} M\right )\right.\label{secondzerdef}\\
&&-2\,r^{9/2}\left (2\,r+3\,M\right ) (K^{(1)})^{2}-4\,r^{5}\rho^
{3} h_{1,r}^{(1)}  K\!K^{(1)}  +4\, K^{(1)} r^{6} K\!K^{(1)}  \rho-4\,r
^{4} h_1^{(1)}  K\!K^{(1)}  M\rho+2\,r^{6} K\!K^{(1)}   H_2^{(1)} \rho
\nonumber\\
&&+6\,r^{5}\rho\,\left (5\,r+9\,M\right ) K\!G^{(1)}   G^{(1)} +4\,r^{5}
\rho\,\left (-2\,r+M\right ) K\!G^{(1)}  H_2^{(1)} -r^{6}\rho\,\left(
-3\,r+4\,M\right ) K\!K^{(1)}  G^{(1)}_r \nonumber\\
&&+r^{6}\rho\,\left (-13\,r+20
\,M\right ) K\!G^{(1)}  G^{(1)}_r -2\,r^{5}\rho\,\left (8\,r+9\,M
\right ) K^{(1)}  K\!G^{(1)} +4\,r^{4}\rho\,\left (-r+5\,M\right )
 K\!G^{(1)}  h_1^{(1)} \nonumber\\
&&-2\,r^{5}\rho\,\left (7\,r+6\,M\right )
 K\!K^{(1)}  G^{(1)} +r^{5}\rho^{5} K\!h^{(1)}_{1,r}  G^{(1)}_r +2\,r^{6
}\rho^{3} K\!G^{(1)}  K^{(1)}_r 
+4\,r^{5}\rho^{3} K\!h_1^{(1)} 
 K^{(1)}_r +4\,r^{5}\rho^{3} K\!K^{(1)}_{r}  h_1^{(1)} \nonumber\\
&&-6\,r^{4}\rho^{3}
 G^{(1)}  K\!h_1^{(1)} 
-2\,r^{7}\rho^{3} K\!G^{(1)}_{r}  K^{(1)}_r -2\,r
^{7}\rho^{3} K\!K^{(1)}_{r}  G^{(1)}_r -2\,r^{3}\rho^{5} K\!h^{(1)}_{1,r} 
 h_1^{(1)} 
-2\,r^{4}\rho^{3} h_1^{(1)}  K\!H_2^{(1)} \nonumber\\
&&+r^{6}\rho^{3}
 G^{(1)}_r  K\!H_2^{(1)} +4\,r^{3}\rho^{5} K\!h_1^{(1)}  h_{1,r}^{(1)} 
-7\,r^{4}\rho^{3} K\!h_1^{(1)}  H_2^{(1)} +2\,\rho^{3}r^{2}\left (7\,M+
22\,r\right ) K\!h_1^{(1)}  h_1^{(1)} \nonumber\\
&&-8\,\sqrt {r}\left (2\,r+3\,M\right )\rho^{4} (h_1^{(1)}) ^{
2}+r^{4}\rho^{3}\left (M-26\,r\right ) K\!h_1^{(1)}  G^{(1)}_r -24\,r^
{5}\rho^{3} K\!G^{(1)}_{r}  h_1^{(1)} +12\,r^{5}\rho^{3} h_{1,r}^{(1)} 
 K\!G^{(1)} \nonumber\\
&&+12\,r^{7}\rho^{3} K\!G^{(1)}_{r} 
 G^{(1)}_r +2\,r^{6}\rho^{3} K\!G^{(1)}_{r}  H_2^{(1)} -4\,r^{9/2}\left(
2\,r+3\,M\right )\rho^{2} G^{(1)}_r  K^{(1)} +8\,r^{5/2}\left (2\,r+3
\,M\right )\rho^{4} h_1^{(1)}  G^{(1)}_r \nonumber\\
&&\left. +8\,r^{5/2}\left (2\,r+3\,M
\right )\rho^{2} h_1^{(1)}  K^{(1)} -2\,r^{9/2}\left (2\,r+3\,M
\right )\rho^{4} (G^{(1)}_r)^{2}\right]
{[7\,r^{9/2}\left (2\,r+3\,M\right )]^{-1}}\nonumber
\end{eqnarray}
where $\rho=\sqrt{r-2M}$, and the time derivative of
the second order, $\ell=2$, Zerilli function 
is given by,
\begin{eqnarray}\label{genchidot}
\chi_{,t} &=&
-4\,r^{15/2}\left (7\,M-6\,r\right )\rho^{6} H_2^{(1)} 
 G^{(1)}_r +16\,Mr^{7}\rho^{5}\left (2\,r+3\,M\right ) K^{(1)} 
 K\!G^{(1)} -4\,r^{17/2}\rho^{10} K\!G^{(1)}_{r}  K\!h^{(1)}_{1,r} \\
&&
+28\, H_2^{(2)} r^{13/2}\left (M+r\right )\rho^{6}
+2\,r^{13/2}\left (18\,M^{2}+17\,rM-8\,r^{2}\right )
\rho^{6} G^{(1)}  K^{(1)}_r
 +4\,r^{13/2}\left (5\,r+9\,M\right )\rho^{6} G^{(1)}  H_2^{(1)} 
\nonumber\\&&
-4\,r^{15/2}\left (M+12\,r\right )\rho^{8} K\!G^{(1)}  K\!h^{(1)}_{1,r} 
+4\,r^{13/2}\left (2\,r+3\,M\right )\rho^{8} G^{(1)}  H_{2,r}^{(1)} 
+4\,r^{15/2}\left (3\,M+50\,r\right )\rho^{8} K\!G^{(1)}_{r}  K\!h_1^{(1)} 
\nonumber\\&&
-4\,r^{17/2}\left (17\,M+15\,r\right )\rho^{6} (K\!G^{(1)})^{2}
+8\,r^{19/2}\left (3\,M-r\right )\rho^{6} K\!G^{(1)}  K\!K^{(1)}_{r} 
+4\,r^{13/2}\left (6\,M-5\,r\right )\rho^{6} G^{(1)}  K^{(1)} 
\nonumber\\&&
-12\,r^{11/2}\left (2\,r+3\,M\right )\rho^{8} h_{1,r}^{(1)}  G^{(1)} 
-48\,r^{21/2}\rho^{8} (K\!G^{(1)}_{r})^{2}
-116\,r^{13/2}\rho^{8} (K\!h_1^{(1)})^{2}
\nonumber\\&&
-28\,r^{15/2}\rho^{6} K^{(2)} 
-12\,r^{19/2}\rho^{6} (K\!K^{(1)})^{2}
+14\,r^{15/2}\rho^{6} (K^{(1)})^{2}
-6\,r^{13/2}\left (12\,M^{2}+19\,rM-6\,r^{2}\right )
\rho^{6} G^{(1)}_r  G^{(1)} 
\nonumber\\&&
-24\,r^{9/2}\left (10\,M-3\,r\right )\rho^{8} h_{1,r}^{(1)}  h_1^{(1)} 
+4\,r^{19/2}\left (2\,M-3\,r\right )\rho^{6} K\!G^{(1)}_{r}  K\!K^{(1)} 
+14\,r^{15/2}M\rho^{6} K^{(2)}_r 
\nonumber\\&&
-12\,r^{7/2}\left (42\,M^{2}-14\,rM-3\,r^{2}\right )\rho^{6} (h_1^{(1)})^{2}
+3\,r^{15/2}\left (54\,M^{2}-30\,rM+r^{2}\right )\rho^{6} (G^{(1)}_r)^{2}
\nonumber\\&&
+2\,r^{17/2}\rho^{8} H_{2,r}^{(1)}  K^{(1)}_r 
-28\,r^{9/2}\left (6\,M^{2}+5\,rM-2\,r^{2}\right )\rho^{6}h_1^{2}
-3\,r^{17/2}\rho^{8} (K^{(1)}_r)^{2}
\nonumber\\&&
-r^{13/2}\left (2\,M-5\,r\right )\rho^{6} (H_2^{(1)})^{2}
-28\,r^{11/2}\left (2\,r+3\,M\right )\rho^{8}h^{1,2}_r
+4\,r^{11/2}\left (4\,M-7\,r\right )\rho^{8} h_{1,r}^{(1)}  H_2^{(1)} 
\nonumber\\&&
+16\,r^{21/2}\rho^{8} K\!G^{(1)}_{r}  K\!K^{(1)}_{r} 
+48\,r^{15/2}\rho^{6} (G^{(1)})^{2}
+16\,r^{8}\rho^{5}\left (2\,r+3\,M\right ) K\!K^{(1)}  K^{(1)} 
\nonumber\\&&
+64\,r^{4}\rho^{9}\left (2\,r+3\,M\right ) K\!h_1^{(1)}  h_1^{(1)} 
+16\,r^{8}\rho^{7}\left (2\,r+3\,M\right ) K\!K^{(1)}  G^{(1)}_r 
-32\,Mr^{5}\rho^{7}\left (2\,r+3\,M\right ) K\!G^{(1)}  h_1^{(1)} 
\nonumber\\&&
+16\,r^{8}\rho^{7}\left (2\,r+3\,M\right ) K\!G^{(1)}_{r}  K^{(1)} 
+2\,r^{15/2}\left (3\,M-r\right )\rho^{8} G^{(1)}_r  H_{2,r}^{(1)} 
+16\,r^{7}M\rho^{7}\left (2\,r+3\,M\right ) K\!G^{(1)}  G^{(1)}_r 
\nonumber\\&&
-32\,r^{6}\rho^{9}\left (2\,r+3\,M\right ) K\!h_1^{(1)}  G^{(1)}_r 
-32\,r^{6}\rho^{7}\left (2\,r+3\,M\right ) K\!K^{(1)}  h_1^{(1)} 
+16\,r^{8}\rho^{9}\left (2\,r+3\,M\right ) K\!G^{(1)}_{r}  G^{(1)}_r 
\nonumber\\&&
-32\,r^{6}\rho^{7}\left (2\,r+3\,M\right ) K\!h_1^{(1)}  K^{(1)} 
-12\,r^{13/2}\left (2\,r+3\,M\right )\left (M-r\right )
\rho^{6}G^{(1)}_rK^{(1)}
 -40\,r^{11/2}\rho^{8} h_1^{(1)}  K^{(1)} 
\nonumber\\&&
+12\,r^{13/2}\left (10\,M-3\,r\right )\rho^{8} G^{(1)}_r  h_{1,r}^{(1)} 
+12\,r^{11/2}\left (8\,M^{2}+11\,rM-7\,r^{2}\right )
\rho^{6} h_1^{(1)} G^{(1)}_r
\nonumber\\&&
+8\,Mr^{9/2}\left (4\,M-7\,r\right )\rho^{6} h_1^{(1)}  H_2^{(1)} 
+36\,r^{9/2}\left (4\,M^{2}-2\,r^{2}+rM\right )\rho^{6} h_1^{(1)}  G^{(1)}
\nonumber\\&&
+4\,r^{13/2}\left (5\,M^{2}+2\,rM-2\,r^{2}\right )
\rho^{6} K\!G^{(1)}K\!h_1^{(1)}
-4\,r^{11/2}\left (3\,M-r\right )\rho^{8} H_{2,r}^{(1)}  h_1^{(1)} 
-12\,r^{19/2}\rho^{8} K\!G^{(1)}_{r}  K\!H_2^{(1)} 
\nonumber\\&&
-4\,r^{11/2}\left (35\,M^{2}-19\,rM+r^{2}\right )\rho^{6} K^{(1)}_r h_1^{(1)} 
+16\,r^{17/2}\rho^{8} K\!h^{(1)}_{1,r}  K\!K^{(1)} 
-2\,r^{15/2}\left (27\,M^{2}-7\,r^{2}\right )\rho^{6} G^{(1)}_r K^{(1)}_r 
\nonumber\\&&
+8\,r^{13/2}\rho^{8} h_{1,r}^{(1)}  K^{(1)} 
+32\,r^{15/2}\rho^{8} K\!h_1^{(1)}  K\!H_2^{(1)} 
-4\,r^{13/2}\left (10\,M-3\,r\right )\rho^{8} h_{1,r}^{(1)}  K^{(1)}_r 
\nonumber\\&&
+16\,r^{15/2}M\rho^{6} K\!h_1^{(1)}  K\!K^{(1)} 
+4\,r^{17/2}\left (M+12\,r\right )\rho^{6} K\!G^{(1)}  K\!H_2^{(1)} 
\nonumber\\&&
 -4\,r^{15/2}\rho^{6} K^{(1)}  H_2^{(1)} 
-14\,r^{13/2}\left (6\,M^{2}-3\,rM-2\,r^{2}\right )\rho^{6} G^{(2)}_r 
+4\,r^{17/2}\left (17\,M+16\,r\right )\rho^{6} K\!G^{(1)}  K\!K^{(1)} 
\nonumber\\&&
-4\,r^{15/2}M\rho^{6} K^{(1)}_r  K^{(1)} 
-8\,r^{13/2}\rho^{10} K\!h^{(1)}_{1,r}  K\!h_1^{(1)} 
-2\,r^{15/2}\rho^{8} H_2^{(1)}  H_{2,r}^{(1)} 
\nonumber\\&&
-4\,r^{19/2}\left (-13\,r+20\,M\right )\rho^{6} K\!G^{(1)}  K\!G^{(1)}_{r} 
-8\,r^{19/2}\rho^{6} K\!H_2^{(1)}  K\!K^{(1)} 
-32\,r^{17/2}\rho^{8} K\!K^{(1)}_{r}  K\!h_1^{(1)} 
\nonumber\\&&
\left.-32\,r^{6}\rho^{9}\left (2\,r
+3\,M\right ) K\!G^{(1)}_{r} h_1^{(1)} \right]
\left[14\,\rho^{4}\left (2\,r+3\,M\right )r^{17/2}\right]^{-1}\nonumber
\end{eqnarray}
where a subscript $r$ denotes differentiation. To arrive at the
expressions in (\ref{secondzerdef}) and (\ref{genchidot}) the second
order Einstein equations have been used to eliminate higher order time
derivatives.  The above expressions were automatically computed with
Maple computer algebra codes. It is impractical to give more details
of their construction in print. The source codes and documentation can
be found in our anonymous ftp server \cite{anon}.

When the explicit 3-geometry and extrinsic curvature of
(\ref{final3geom})--(\ref{finalext}) are put into the expressions
of (\ref{secondzerdef}) and (\ref{genchidot}), we arrive at the
following initial data for the first and second order Zerilli
equations,
\begin{eqnarray}
\psi_{t=0} &=& 
{\displaystyle \frac {8}{3}}{\displaystyle 
\frac {{M}{L}^{2} \left(  5\sqrt {{r} - 2{M}} + 7
\sqrt {{r}}   \right) {r}}{ \left(  \sqrt {{r}} + 
\sqrt {{r} - 2{M}}   \right) ^{5}(2{r} + 3{M})}}\label{psit0}\\
\dot{\psi}_{t=0} &=& 
{\displaystyle \frac {\sqrt{{r} - 2{M}}
{P}{L}(8
{r} + 6{M})}{{r}^{5/2}(2{r} + 3{M})}}\\
\chi_{t=0} &=& - {\displaystyle \frac {512}{7}}
{\displaystyle \frac {{M}^{2}{L}^{4}}{ \left(  \sqrt {{r}}
 + \sqrt {{r} - 2{M}}   \right) ^{10}{r}}} +
{\displaystyle \frac {16}{7}}{\displaystyle 
\frac { \left(  9\sqrt {{r}} + 
17\sqrt {{r} - 2{M}}   \right) \sqrt {r - 2{M}}{M}{P}{L}^{3}}{
 \left(  \sqrt {{r}} + \sqrt {{r} - 2{M}}   \right) ^{5
}
(2{r}
 + 3{M}){r}^{5/2}}}\\
\dot{\chi}_{t=0} &=&
{\frac {64\,{M}^{2}\left (10\,r-10\,M+38\,\sqrt {r}\sqrt {r-2\,M}
\right )\sqrt {r-2\,M}{L}^{4}}{7\,\left (\sqrt {r}+\sqrt {r-2\,M}
\right )^{10}\left (4\,r+6\,M\right ){r}^{5/2}}} - \nonumber \\
&-& \,{\displaystyle \frac {64}{7}}\,
{\displaystyle \frac {(\,4\,{r} + 14\,{M}\,)\sqrt{r-2M}\,{M}\,{P}\,{L}^{3}}{
 \left( \! \,\sqrt {{r}} + \sqrt {{r} - 2\,{M}}\, \!  \right) ^{5
}\,{r}^{5}}}
- 64\,{\displaystyle \frac {{M}\,
 \left( \! \,5\,\sqrt {{r}} - 3\,\sqrt {{r} - 2\,{M}}\, \! 
 \right) \,\sqrt {{r} - 2\,{M}}\,{L}^{2}\,{P}^{2}\,{\it q2}}{
 \left( \! \,\sqrt {{r}} + \sqrt {{r} - 2\,{M}}\, \!  \right) ^{5
}\,(\,2\,{r} + 3\,{M}\,)\,{r}^{5/2}}} \nonumber\\
 &&+ 
{\displaystyle \frac {16}{35}} \sqrt {{r} - 2\,{M}}\,
{L}^{2}\,{P}^{2} \left( 1750 M^4-9849rM^3+2331r^2M^2+7182r^3M-2892 r^4 \right. \nonumber \\
&&
\left.
-\sqrt{r} \sqrt{r-2M} (3148 r^3-4130r^2M-4935rM^2+4375M^3) 
 \right) \left/ \right. \! \!  \left( \! \,(\,2\,{r} + 3\,{M}\,)\, \left( \! \,
\sqrt {{r}} + \sqrt {{r} - 2\,{M}}\, \!  \right) ^{5}\,{r}^{6}\,
 \!  \right)  .\label{chidott0}
\end{eqnarray}

We are now ready to evolve the initial data and compute waveforms and
radiated powers.

\section{Evolution}
\label{Evolution}
\subsection{The Zerilli equations}
The initial data generated in the previous section is now fed to the
first and second order Zerilli equations
\begin{eqnarray}
-{\partial^2 \psi \over \partial t^2} 
+{\partial^2 \psi \over \partial r_*^2} 
+V(r) \psi &=& 0\label{firstorderzeq}\\
-{\partial^2 \chi \over \partial t^2} 
+{\partial^2 \chi \over \partial r_*^2} 
+V(r) \chi &=& {\cal S}\label{secorderzeq}
\end{eqnarray}
where $r_*$ is the usual ``tortoise'' coordinate covering the exterior
of the black hole,
\begin{equation}
r_* = r+2 M \ln({r/2M} -1)
\end{equation}
so the horizon is at $r_*=-\infty$ and spatial infinity at
$r_*=\infty$, and the potential and source terms in the
Zerilli\cite{zerilli} equations are given by,
\begin{eqnarray}
V(r) &=&
\left(1-\frac{2M}{r}\right)
  \left\{\frac{4 r^2}{\Delta^2}
  \left[\frac{72 M^3}{r^5}-\frac{12 M}{r^3} (l-1)(l+2)
  \left(1-\frac{3M}{r} \right)\right]
 + \frac{2(l-1)(l+2)l(l+1)}{r\Delta}\right\}\label{zpot}\\
{\cal S} &=& 
  {12 \over 7} {\mu^3 \over \Delta} \left[ -{12
(r^2+Mr+M^2)^2 \over r^4\mu^2\Delta} \left(\psi,_t\right)^2 -4
{(2r^3+4r^2M+9rM^2+6M^3) \over r^6\Delta} \psi \psi,_{rr}  \right.
\nonumber \\
& &  +{(112r^5+480r^4M+692r^3M^2+762r^2M^3+441rM^4+144M^5) \over
r^5\mu^2\Delta^3} \psi \psi,_t - {1 \over 3r^2} \psi,_t \psi,_{rrr}
\nonumber \\
& &   +{18r^3-4r^2M-33rM^2-48M^3 \over 3 r^4\mu^2\Delta} \psi,_r
\psi,_t + {12r^3+36r^2M+59rM^2+90M^3 \over 3 r^6\mu} \left(\psi,_r
\right)^2 \nonumber \\
& &\! +\! 12 {(2r^5 + 9r^4M +6r^3M^2\!-\!2r^2M^3\!-\!15rM^4-15M^5) \over
r^8\mu^2\Delta} \psi^2 \!-\!4 {(r^2+rM+M^2) \over r^3\mu^2} \psi,_t\!
\psi,_{tr} \nonumber \\
& & -2 {(32r^5+88r^4M+296r^3M^2+510r^2M^3+561rM^4+270M^5) \over
r^7\mu\Delta^2} \psi \psi,_r  + { 1 \over 3r^2 } \psi,_r \psi,_{trr}
\nonumber \\
& & - {2r^2-M^2 \over r^3\mu\Delta} \psi,_t \psi,_{rr}
 +{8r^2+12rM+7M^2 \over r^4\mu\Delta} \psi \psi,_{tr}  +{3r-7M \over
 3r^3\mu} \psi,_r \psi,_{tr} - {M \over r^3\Delta} \psi \psi,_{trr}
 \nonumber \\
& &  + {4(3r^2+5rM+6M^2) \over 3r^5} \psi,_r \psi,_{rr} \left.  +
{\mu\Delta \over 3 r^4} \left( \psi,_{rr} \right)^2 - {2r+3M \over
3r^2\mu} \left( \psi,_{tr} \right)^2 \right]
\end{eqnarray}
where $\Delta\equiv(\lambda r+3 M)$ with $\lambda = (\ell-1)(\ell+2)/2$
and $\mu\equiv(r-2 M)$. The potential
is given for general $\ell$ in (\ref{zpot}) but we will use it only
for $\ell=2$.

As can be seen, equations (\ref{firstorderzeq}) and
(\ref{secorderzeq}) have the same form, including the same potentials,
but the second order equation has a source term that is quadratic in
the first order Zerilli\cite{zerilli} function and its time
derivatives.  We have written a fortran code to evolve these equations
by a simple leapfrog algorithm. Convergence to second order was
checked and special care was taken to avoid noise from the high
derivative order of the source term.

To find the gravitational waveforms and power a transformation must be
made to a gauge that is asymptotically flat to first and second order.
The details of this process were discussed in GNPP and will not be
repeated here.  The result  is that
the transverse-traceless perturbations, in the asymptotically flat
gauge, correct to second order, are encoded in the quantity
\begin{equation}\label{waveform}
  f(r,t)=\frac{\partial\psi}{\partial t}+\left[\chi
+\frac{1}{7}\frac{\partial}{\partial t}
\left(\psi\frac{\partial\psi}{\partial t}\right)\right]\ ,
\end{equation}
(where it is understood that all quantities are $\ell=2$) and this is
the quantity we shall plot below when we give waveforms of the
outgoing gravitational radiation. The first order part of the
radiation is given by the leading term in (\ref{waveform}); the terms
in square brackets are second order.  From the Landau-Lifschitz
pseudo-tensor in the asymptotically flat gauge (as discussed in GNPP)
we find that the radiated power is
\begin{equation}\label{power}
{\rm Power}= 
{3\over 10} \left\{ {\partial \psi \over \partial t} +
\left[
 \chi  +
{1 \over 7} {\partial \over \partial t} \left( \psi 
{\partial \psi \over \partial
t}\right)\right]\right\}^2. 
\end{equation}
(Note that the perturbation parameter $\epsilon$ that appeared in
\cite{GlNiPrPucqg} is now incorporated into the definition of the
Zerilli functions we have used in the paper, as can be seen in
formulas (\ref{psit0}-\ref{chidott0}). We have also directly computed
the ``renormalized'' second order Zerilli function in
(\ref{secondzerdef})).

Before we move on to present our results and compare with the
numerical relativity simulations of the Potsdam/NCSA/WashU group (see
BAABRPS), it is worth pointing out, again, that the numerical
relativity simulations are for ``symmetrized'' initial data, in which
an infinite number of ``image charges'' is used to construct initial
data representing two throats connecting two isometric asymptotically
flat universes. By contrast, the problem we are solving corresponds
to three asymptotically flat universes.  In the limit of zero momentum
the numerical simulations correspond to the 1960 Misner\cite{Mi}
initial data, and our results correspond to the Brill-Lindquist
\cite{BrLi} initial data.

For the range of separations we are going to discuss the discrepancies
between these two types of data are insignificant. (Although we are
working in the ``close--limit,'' we will consider sets of data far
apart enough to make the extra terms arising from symmetrization very
small), but since the problem is a multi-parametric one, it is not
obvious that this is true in all the ranges of parameters we will be
discussing. More careful studies will be needed if one wants higher
accuracies than the ones we are going to discuss here. We have also
modified the Potsdam/NCSA/WashU code to run for unsymmetrized data,
and for limited tests the results agree very well with the symmetrized
ones in the range we are considering. This situation arose due to
historical reasons: the numerical code was written before our work
with non-symmetrized boundary conditions, whereas perturbation theory
becomes very cumbersome if one starts carrying around the extra terms
due to symmetrization.

One particular problem that one faces when comparing Brill--Lindquist
(unsymmetrized) and symmetrized data sets is that the sets are
parameterized in different ways.  There is therefore ambiguity in how
to compare the results.  Abrahams and Price \cite{AbPr} have discussed
this in some detail, and show that there are different identifications
one can take that yield sensible results along a good range of
parameters, so we will not repeat the discussion here. We just state
the convention we are following: For one of our results with momentum
parameter $P$ and throat separation $L$, we compare a numerical
relativity result with the same ADM mass and same numerical value of
the momentum parameter, and with a separation parameter $\mu_0$ given
by
\begin{equation}
L = 2 \sqrt{4 \kappa_2(\mu_0)}  \ .
\end{equation}
Here $\mu_0$ is 
a parameter originally introduced by Misner \cite{Mi}
that is commonly used to parameterize symmetrized binary black hole
initial data sets, and 

\begin{eqnarray}\label{kapdef}                                   
\kappa_\ell(\mu_0) &\equiv& \frac{1}{\left(4\Sigma_1\right)^{\ell+1}} 
\sum_{n=1}^\infty         \frac{(\coth{n\mu_0})^\ell}{\sinh{n\mu_0}}\\
\Sigma_1&\equiv&\sum_{1}^\infty\frac{1}{\sinh{n\mu_0}}\ .
\end{eqnarray}
With these choices, the radiated waveforms agree very well when $P=0$.
Notice that the discussion of Abrahams and Price \cite{AbPr} is only
for the $P=0$ case. The ``best'' identification between symmetrized
and unsymmetrized data could probably be a $P$-dependent notion. We
will ignore this issue here, but it clearly requires further study.

\subsection{Fixing $t=0$}

In the formalism of GNPP we chose to fix the coordinates by requiring
that the metric be in the Regge-Wheeler\cite{ReWh} gauge to first and
second order. This can always be done, but it turns out that the
coordinates are not quite uniquely fixed.  The problem is quite
generic and it has to do with how perturbation theory handles time
translations in situations where the background spacetime is
time-translation invariant. Consider an exact quantity
$f(r,t)$ approximated by a perturbative series
expansion,
\begin{equation}
f(r,t)= f^{(0)}(r) 
+ \epsilon f^{(1)}(r,t)
+ \epsilon^2 f^{(2)}(r,t) +\ldots,
\end{equation}
and perform now a first order gauge transformation corresponding to a
pure time translation,
\begin{equation}\label{ttrans}
t\rightarrow t'=t+\epsilon c ,
\end{equation}
with $c$ a constant, independent of $t$ and $r$. 
Replacing $t'$ by the above expression (and
noticing that $dt=dt'$), we get
\begin{eqnarray}
f(r,t')&=& f^{(0)}(r) 
+ \epsilon f^{(1)}(r,t')
+ \epsilon^2 f^{(2)}(r,t')
+O(\epsilon^3)\nonumber\\ &=&
f^{(0)}(r) + \epsilon f^{(1)}(r,t)
+ \epsilon^2 (c\dot{f}^{(1)}(r,t) +f^{(2)}(r,t)) +
O(\epsilon^3).
\end{eqnarray}
So we see that the ``second order term'' in the expansion of the
metric depends on the origin chosen for time. If one starts with
perturbations in the Regge-Wheeler gauge, a transformation of type
(\ref{ttrans}) leaves the perturbations in the Regge-Wheeler gauge,
but the second-order metric is changed, and in fact depends on an {\em
  arbitrary} constant $c$.  This indicates that a comparison of
quantities to second order in perturbation theory around stationary
backgrounds can be quite misleading: the same metric can have very
different second order terms depending on the origin of time chosen.
Worse, these terms can be quite large, and are completely artificial.

It is interesting to notice that if one computes the radiated energies
using the formula we discussed previously (\ref{power}), the results
are unchanged ---as expected--- by time translations (the additional
term turns out to be a total derivative that does not affect the
computation of energies).  But we want to go beyond giving
perturbative results for radiated energy. We want also to compare
perturbative waveforms with those of numerical relativity. Since these
waveforms are second-order correct quantities given as a function of
time at a particular ``observation'' radius, we must be sure that we
are using the same zero of time for both waveforms, that from
perturbation theory and that computed with numerical relativity.

Fortunately, it is not difficult to eliminate the time-shift ambiguity
in the metric. To do this we separate the waveform $f(r,t)$ given in
(\ref{waveform}) into first and second order parts $f^{(1)}$ and
$f^{(2)}$ and we construct the quantity
\begin{equation}\label{c0eq}
c_0 = \frac{\int_{-\infty}^{\infty} dt 
\dot{f}^{(1)}(r,t) f^{(2)}(r,t)}
{\int_{-\infty}^{\infty} dt \left[\dot{f}^{(1)}(r,t)\right]^2}\ .
\end{equation}
We then perform the time translation $t\rightarrow t+c_0$, arriving
at the ``physical'' value for the second order waveform
\begin{equation}
f^{(2)}_{\rm phys}(r,t) = f^{(2)}(r,t)- c_0 \dot{f}^{(1)}(r,t)\ .
\end{equation}
Equivalently, we adjust the zero of time, and hence $f^{(2)}(r,t)$
until the integral in the numerator of (\ref{c0eq}) vanishes.  The
same coordinate fixing must be done to the numerically computed
waveform $f_{\rm num}$. To do this we define $f^{(2)}_{\rm num}$ to be
$f_{\rm num}-f^{(1)}$. We then adjust the zero of time so that the
integral of $f^{(2)}_{\rm num}\dot{f}^{(1)}$ vanishes.

These observations about time-shifts are also true in the
time--symmetric case. We have recomputed the results of
\cite{GlNiPrPuprl} with the zero of time fixed as above and have found
that the results are changed by less than $1\%$.  For boosted black
holes, on the other hand, this time fixing is crucial for seeing the
high accuracy agreement of the perturbative and numerical relativity
results.

\subsection{Results: radiated energies}

\begin{figure}[h]
  \centerline{\psfig{figure=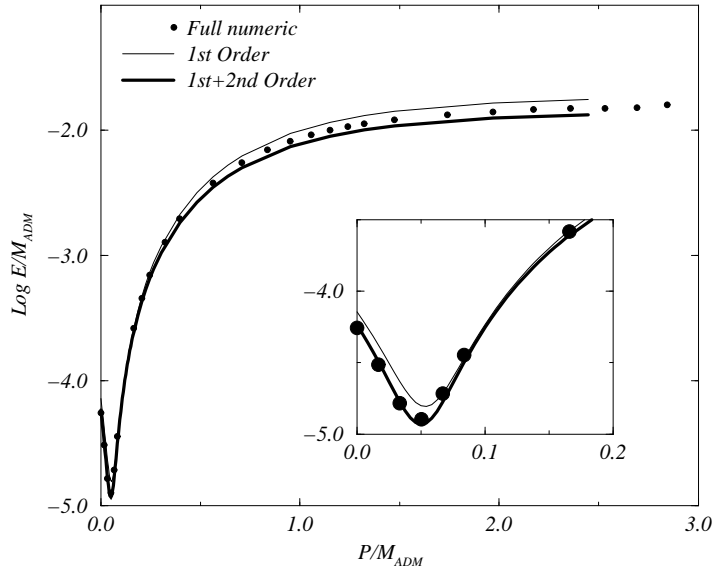,height=3.5in}}
\caption{Radiated energy in head-on black hole collisions as function
of the momentum for a separation of $\mu_0=1.5$, $L/(0.5 M_{ADM})=
4.2$.  Depicted are the close-slow approximation and the full
numerical results of the Potsdam/NCSA/WashU group.  Even for large
values of the momentum, the first order results overshoot and the
first plus second order undershoot the numerical results by only
$20\%$. The inset shows the ``dip'' region.}
\label{fig1}
\end{figure}

We start to summarize our results by computing the radiated energy as
a function of momentum for head-on collisions of black holes released
from a separation of $\mu_0=1.5$, $L/(0.5 M_{ADM})= 4.2$. The results
are depicted in Fig.\,\ref{fig1}.

The figure shows the characteristic ``dip'' at low values of the
momentum that was first noticed in BAABRPS. An important difference
between that paper and the present results is that here, as explained
in Sec.\,IIC, we are normalizing both the numerical and the
perturbative results using the same ADM mass. This leads to a much
better agreement for large values of the momentum than that observed
in BAABRPS. As an example of the size of the difference, for
$P/M_{ADM}=1$, $P/(2m)\approx 3$ and for $P/M_{ADM}=3$, $P/(2m)\approx
15$.

A remarkable fact is that first order perturbation theory agrees very
well even for large values of the momentum, and second order
perturbation theory confirms this fact.  This at first seems puzzling
since our initial data was obtained through a ``slow'' approximation
in which the momentum was assumed to be small. However, as was
observed in BAABRPS, for large values of the momentum the initial data
is ``momentum dominated'', meaning that the extrinsic curvature
completely dominates the initial data. Therefore the errors made in
computing the conformal factor via the slow approximation become less
relevant than might be supposed.

\begin{figure}
\centerline{\psfig{figure=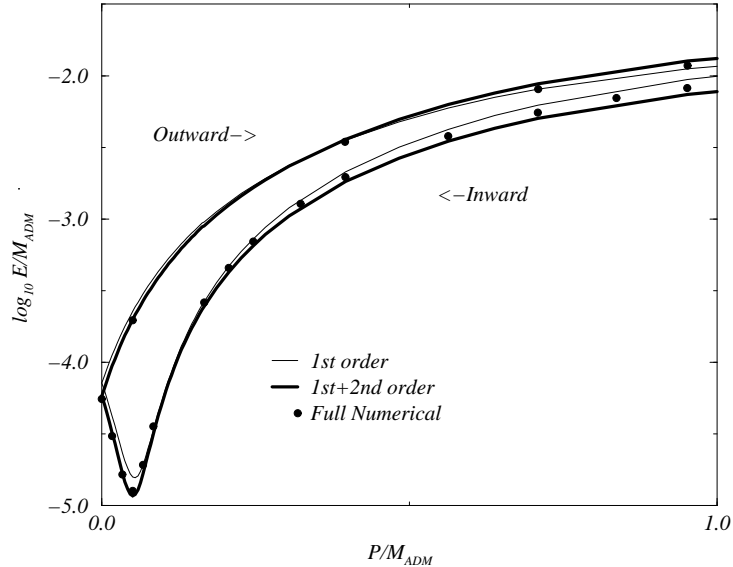,height=3.5in}}
\caption{We depict energies for black hole collisions in which black
  holes were initially boosted towards each other and also for the
  case in which they were boosted away from each other.  The ``dip''
  effect is only present when the black holes are moving towards each
  other. The approximation given by first order perturbation theory is
  slightly worse in that case, since leading terms in the calculation
  are cancelling each other to produce the suppression. We also see
  the cancellation of second order contributions that takes place for
  the outward pointing momentum case. There, first order results
  undershoot the results for small values of $P$ and overshoot it
  later, the second order corrections vanishing at the point of
  crossing.}
\label{fig2}
\end{figure}

The overall picture of the energy therefore is very encouraging, the
approximations presented seem to be working even beyond their expected
realm, and second order perturbation theory is capable of tracking
this fact, playing the expected role of ``error bars.''  This approach
is not without pitfalls, however. In order to illustrate these, we
turn to Fig.\,\ref{fig2}, which shows a close up look at the energy
picture and also includes results for black holes initially boosted
{\em away from each other}.

The first thing we notice is, that for black holes boosted away from
each other there is, as expected, no ``dip'' in the energy. The dip is
a first order effect that is due to a cancellation between terms that
are momentum independent and terms that are linear in momentum. The
cancellation turns to addition in the case of negative (outwards) $P$.
We also see that first order calculations are less accurate at the dip
than at higher values of the momentum. This is somewhat puzzling since
our approximation should work better the smaller the momentum.  What
seems to be happening is that first order theory does not accurately
reproduce the higher order terms that make important contributions to
the energy after the leading terms cancel in order to produce the dip.
This is confirmed by the fact that first plus second order results are
indeed very accurate at the dip.

An instructive feature of these results is that for black holes
boosted away from each other a cancellation of the second order terms
takes place around $P/(2m)=0.9$, $P/M_{ADM}=0.36$. Clearly one cannot
regard second order perturbation theory as giving error bars when it
is cancelling out.  Moreover, it shows that second order results {\em
  beyond} that value of $P$ can only be taken as rough indicators. We
will return to this cancellation in somewhat more detail in connection
with waveforms.

Another issue to be mentioned is how crucial it is to have chosen the
$ADM$ mass of the initial slice as the mass of the background
spacetime used in the perturbative calculations. Our previous
(first-order) work on boosted black holes used the ``bare'' mass (ADM
mass for $P=0$) for the background. This is quite visible if one
compares Fig.\,\ref{fig1} with Fig.\,2 of BAABRPS.  In the latter,
first order perturbation results appeared to disagree with the
numerical results by over an order of magnitude for $P/M_{ADM}=3$.
That was entirely due to the poor choice of background mass. In the
present paper, using the numerically computed ADM mass of the initial
data, we see that first (and second) order results differ by only
$20\%$ from the numerical results at $P/M_{ADM}=3$.

\subsection{Results: waveforms}

Let us now turn to the examination of waveforms. The numerical code of
the Potsdam/NCSA/WashU group extracts waveforms at slightly different
values of the radial variable for varying $P$'s. We took this effect
into account and extracted perturbative waveforms at the same radii as
was used for the numerical relativity work.  In all cases the full
numerical code has a very limited range of spacetime covered in the
evolution. This forces the extraction to be done in a rather small
range of radii around $20 M_{ADM}$ or so. With perturbation
calculations we could have extracted much further away, but we
performed the extraction at exactly the same radius as those used by
the numerical code. Waveforms were observed to change shape rather
significantly from one extraction radius to another even in such a
close range, but we observed that as long as we extracted the
perturbative waveform at the same radius as the full numerical result
(as opposed to, say, extracting farther out and then shifting the
result back) the agreement was roughly independent of extraction
radius.  However, this starts to hint at a main problem in comparing
waveforms: one needs not only to match amplitudes but it is also
crucial to match the phase, at least if one is interested in high
accuracy. The phase is determined by, among other things, the
extraction radius. Determining the extraction radius, in turn,
requires knowing the ADM mass (since one measures radii in units of
ADM mass). Our full numerical code for computing the ADM mass, in its
present implementation, is accurate to a few percent. (This could be
made better with more computer power than what is presently available
to us; the runs we made had 300 radial zones and 30 angular zones.)
This limits the accuracy with which we know the ADM mass, and hence
the accuracy with which we can determine the phases.  The technique,
discussed in Sec.\,IIIB, of fixing the zero of time is helpful in
giving an objective way of comparing phases.

Let us turn to the results. We present, below, the results for the
waveform $f(r,t)$ as defined in (\ref{waveform}).  This is directly
comparable (up to a time derivative) with the output of the full
numerical relativity code, which outputs a Zerilli\cite{zerilli}
function via the radiation extraction technique of assuming that the
spacetime is a perturbation of Schwarzschild and reading off the
perturbations from the full numerical results.

\begin{figure}[h]
\centerline{\psfig{figure=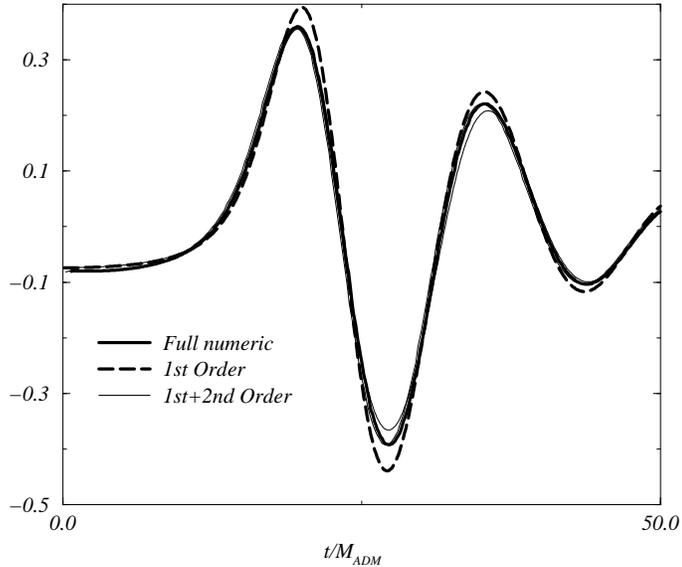,height=3.5in}}
\caption{Comparison of waveforms for large values of the momentum,
  $P/(2m)=14$, $P/M_{ADM} = 2.44$. There is good overall agreement,
  but there is some slight disagreement in the details of the
  waveforms. As one expects for large values of the parameters, second
  order perturbations can at most be regarded as an estimator of
  error, rather than a way to improve the accuracy of the waveforms.
  It is still remarkable that perturbation theory would work well for
  such a large value of the parameter $P$.}
\label{fig3a}
\end{figure}

\begin{figure}
\centerline{\psfig{figure=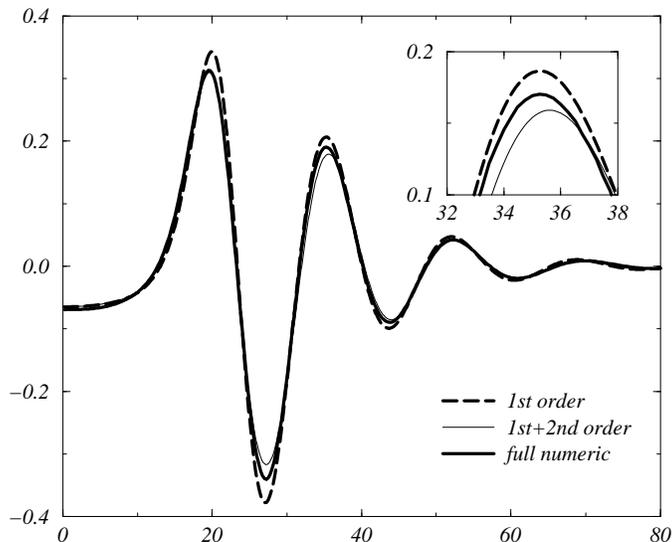,height=3.5in}}
\caption{Comparison of waveforms for intermediate values of the
  momentum, $P/M_{ADM}\sim 1.32$. Here one can see that second order
  theory not only acts as an ``error bar,'' but when added to the
  first order calculation, actually allows a reasonably accurate
  prediction of the waveforms.}
\label{fig3}
\end{figure}

Our presentation of waveform comparisons starts with the most
disfavorable cases and moves to more favorable ones. Figure
\ref{fig3a} shows the comparison of waveforms for $P/(2m)=15$,
$P/M_{ADM}\sim 2.44$ and Fig.\,\ref{fig3} corresponds to $P/(2m)=5$,
$P/M_{ADM}\sim 1.32$. As we see, there is very good overall agreement.
Notice that (taking into account the ``time-shift'' gauge fixing
discussed above) our procedure in the end has {\em no free parameter},
i.e., phases and amplitudes are predetermined in all cases, which
makes the agreement more remarkable. If one looks carefully at the
curves in the inset, which enlarges the region around the second
positive peak, one sees that there are slight phase and amplitude
disagreements. First order results tends to overshoot the waveform,
whereas adding the second order correction tends to undershoot. There
are slight differences in shapes as well. For large values of the
momentum, we can take second order predictions as ``error bars'' only.
However, for intermediate values, it is quite clear that first plus
second order calculations offer a very accurate prediction of the
waveforms.

The reader should exercise care when comparing the results for
waveforms with those of energies. This is due to a peculiarity of the
formula for the radiated power (\ref{power}). As discussed in GNPP,
the square that appears in (\ref{power}) involves terms that are of
``third order'' in perturbation theory. Therefore, to keep things
consistent, when squaring the expression in curly braces, we only keep
the mixed term and omit the term that is the square of the second
order part of $f$. As a consequence, the second order correction for
the radiated energy depends mostly on correlations of phases of the
first and second order waveforms rather than on their amplitudes. For
instance, for the case we are studying $P/(2m)=5$, $P/M_{ADM}\sim
1.32$, the second order waveforms are only slightly smaller than the
numerical ones, but the computed energy is $12\%$ lower.

\begin{figure}
\centerline{\psfig{figure=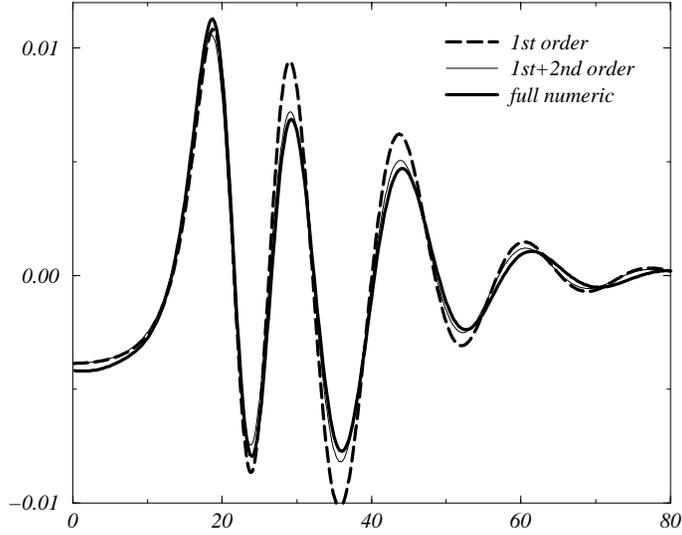,height=3.5in}}
\caption{Waveforms in the ``dip'' region, $P/M_{ADM}\sim 0.05$, for 
  inward boosted black holes (the case where there is a ``dip'' in the
  energy). Second order corrections improve the accuracy in the
  amplitudes significantly, but first order calculations exhibit an
  erratic behavior. Since the energy arises as a correlation between
  both waveforms, this translates itself into a rather large relative
  error in the perturbative computation of the energy.  Notice that
  waveforms for this particular value of the momentum differ markedly
  from waveforms for other values, the ``enveloping'' amplitude curve
  decreasing much more slowly, allowing several oscillations to be
  visible.}
\label{fig5}
\end{figure}

\begin{figure}
\centerline{\psfig{figure=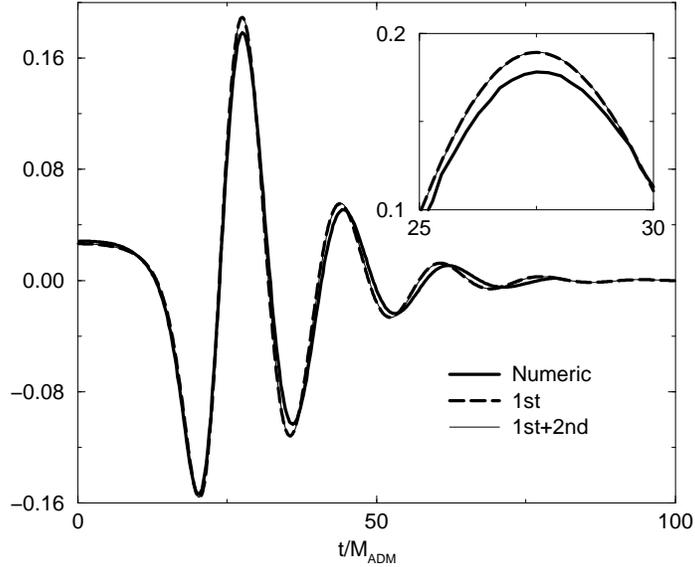,height=3.5in}}
\caption{Waveforms for  $P/(2m)=-0.9$, $P/M_{ADM}\sim -0.36$, i.e., for 
outward boosted black holes (the case where there is no ``dip'' in the
energy). Second order corrections cancel out and therefore are not
reliable as ``error estimators'' nor to improve the accuracy
of first order calculations. (The first and first plus second order 
curves are both plotted but are indistinguishable).}
\label{fig6}
\end{figure}

We now turn our attention to the area of the dip, $P/M_{ADM}\sim
0.05$, $P/(2m)=0.12$. In Fig.\,\ref{fig5} we show the waveforms for
the inward boosted case (the case with a dip in the energy). We see
that second order corrections improve the accuracy markedly. Clearly
there are strange effects taking place for this value of the
parameter.  In particular, it should be noticed how first order theory
overshoots the waveforms rather significantly in the second and third
positive peak of the waveform, but not in the first one. In view of
the fact that the energy is given by the correlation of the first and
second order waveforms, those discrepancies in the first order
waveforms would seem to be responsible for the large relative error in
the calculation for the energy, even to second order. This is so, in
spite of the fact that second order calculations yield very accurate
waveforms.
Figure \ref{fig6} shows the case of $P/(2m)=-0.9$ (holes moving
initially apart).  As could be predicted from the energy plot, a
cancellation of the second order terms is taking place.  In this case,
therefore, one cannot regard second order corrections as ``error
bars,'' since it is clear that higher order terms are important. It is
worthwhile pointing out that the cancellation is highly nontrivial,
the initial data having the same amplitude for both inward and outward
momenta.  The cancellation takes place in the evolution, with the
source terms of the second order Zerilli\cite{zerilli} equation
playing a significant role.
A simple way to understand the cancellation is to break up the
evolution into three separate Zerilli\cite{zerilli} equations with
three different initial sources, proportional to $L^4$, $PL^3$, and
$P^2 L^2$ respectively. What one sees is that the cancellation occurs
between the $PL^3$ term and the other two, and clearly depends on the
sign of $P$ (for our simulations negative $P$ is outward pointing).
One can then infer that there is a curve of cancellations in the
$P$,$L$ parameter space that isolates a region in parameter space
where second order perturbation theory does not help. One cannot reach
points in that region unless one changes the relative counting of
powers of $L$ and $P$ in perturbation theory. A further study of this
issue could therefore yield interesting results.

\section{Discussion}

We have seen that the use of combined first and second order
perturbation theory can give excellent results for waveforms and
energies of radiation emitted in the head on collision of two equal
mass, initially boosted, black holes. The results show, however, that
there are some subtleties, not previously appreciated, in the use of
higher order perturbation theory and in the comparison with results
from numerical relativity. The following points deserve attention,
especially in connection with the application of higher order
perturbation theory to further problems.

a) The comparison of perturbation results and numerical relativity
results has pitfalls when comparisons are made between problems that
are not identical.  In our case we compared our perturbation result
for an ``unsymmetrized'' (Brill-Lindquist\cite{BrLi} type) initial
data, with numerical relativity results for ``symmetrized''
(Misner\cite{Mi} type) initial data. Had we been comparing with
unsymmetrized initial data, the parameters $m,L,P$ for the data sets
would have had identical meaning. Since the data sets were not
identical, a mapping of one parameter set to the other had to be
imposed. One degree of freedom in this mapping was subsumed in the
choice to compare cases of equal ADM mass, but the remaining element
of choice in the mapping is a source of uncertainty in the high
accuracy comparisons we are making. (We emphasize that the choice of
mapping was made before any results were considered; there was no
``fine tuning'' to improve the comparison.  The excellent agreement
between the numerical and perturbative results then must be considered
to be, among other things, an indication that there is no great
sensitivity to the manner in which this mapping of parameters is
done.)

b) There is no unique result that is correct to second order.
Different ways in which details are handled will produce results that
are the same to second order, but differ at higher order. These
different results can have different ranges of validity and can
exhibit different accuracy when compared with numerical work near the
limit of validity.  One example of this feature of perturbation theory
is the dependence on parameterization\cite{AbPr}. In our perturbative
results we have seen another simple example: The second-order correct
waveform consists of a first order and a second order piece. When
radiated energy is computed by squaring this waveform one can choose
simply to take the square, or to truncate the result and omit the
fourth order contribution arising from the square of the second order
contribution to the waveform. (We have made the latter
``conservative'' choice.) Both results, of course, are equally
justifiable for the order of perturbation theory we are doing, but the
results are noticeably different.

c) In the present paper we have seen a particularly interesting
example of the importance of higher order terms and the detailed way
in which perturbation theory is applied.  To make the comparison
between symmetrized and unsymmetrized initial data we found that it is
important to compare cases of equal ADM mass, but the ADM mass (for
fixed $m$) varies quickly with increasing initial momentum. If one
computes this momentum dependence perturbatively the agreement of
perturbation theory and numerical relativity is limited. With the ADM
mass computed exactly (i.e., numerically) the agreement is greatly
improved. This suggests that an {\it a priori} physical understanding
of the dependence on the perturbations can be a very useful guide to
an efficient perturbation scheme.

d) In addition to the numerical computation of ADM mass, another
useful new technical detail was developed in the present work.  A
method was found of fixing the zero of time in the same manner for
both perturbative and numerical waveforms. This fixing of zero had not
been important in previous perturbation studies, but was crucial to
comparison of waveforms for initially boosted holes.

e) Perturbation analysis in the present paper was carried out for both
small separation and small momentum (``the close slow limit''). This
makes it particularly difficult to unravel the sources of disagreement
with numerical results when anomalous cancellations (like the ``dip'')
occur. Although perturbation results end up in excellent agreement
with numerical relativity results, a perturbative analysis based on
small $L$, but without small $P$ (especially if it could be compared
with numerical results for unsymmetrized data), might be useful in
improving our understanding of the nature of errors.

f) The current state of the art of numerical relativity presents
limitations, both in accuracy and in range of simulations of the
codes.  As a consequence, we were limited to comparing waveforms which
are not really in the radiation zone. This is a dangerous exercise
when it comes to second order perturbation theory. In particular, the
formula for the radiated power (from which we extracted the concept of
second order waveform) assumes that one is in the radiation zone. This
is true also of the extraction techniques used in the numerical codes
to produce a Zerilli\cite{zerilli} function as output. In short: with
the current limitations we cannot rule out that the discrepancies we
see in waveforms and energies might be within the error margins of the
numerical results.

A general conclusion of this work is that the synergy between
numerical results and perturbative calculations will probably be one
of the major tools that we will have to use to address with any
accuracy the problem of the collision of two black holes in general
relativity. We see this taking place right now.

\acknowledgements We wish to thank Peter Anninos and Steve Brandt for
help in providing the full numerical results from the NCSA group, and
for allowing us to use the Potsdam/NCSA/WashU code.  We are grateful
to John Baker for several insights concerning the normalization with
the ADM mass. This work was supported in part by grants
NSF-INT-9512894, NSF-PHY-9423950, NSF-PHY-9507719, by funds of the
University of C\'ordoba, the University of Utah, the Pennsylvania
State University and its Office for Minority Faculty Development, and
the Eberly Family Research Fund at Penn State. We also acknowledge
support of CONICET and CONICOR (Argentina). JP also acknowledges
support from the Alfred P. Sloan Foundation. Part of this work was
done while CON was visiting Penn State with support from CONICET
(Argentina).  RJG is a member of CONICET (Argentina)

\end{document}